\documentstyle[aasms4]{article}
\input epsf
\begin{document}
\title{Are Low Surface Brightness Discs Young?}

\author{Paolo Padoan}
\affil{Theoretical Astrophysics Center, Juliane Maries Vej 30, DK-2100 Copenhagen, DK}
\author{Raul Jimenez}
\affil{Royal Observatory, Blackford Hill EH9-3HJ, Edinburgh, UK}
\author{Vincenzo Antonuccio-Delogu}
\affil{Osservatorio Astrofisico di Catania,  
Citt\`{a} Universitaria - Viale A. Doria 6,
95125 Catania, IT}
\authoremail{padoan@tac.dk}
\authoremail{raul@roe.ac.uk}

\begin{abstract}

We reconsider the problem of the age of the stellar discs of
late-type Low Surface Brightness (LSB) galaxies by making
use of a new IMF recently derived from numerical
fluid dynamical simulations (Padoan, 
Nordlund and Jones, 1997). While a Miller-Scalo IMF cannot
adequately describe the photometric properties
of LSBs, when we apply the new Padoan et al. (1997) IMF to a 
simple 
exponential disc model with parameters appropriate to LSBs, 
we get excellent fits of the colors and color gradients. We then
conclude that: a) The star formation history of LSB disc galaxies can be described
 by an initial burst of a few times $10^7 \rm{yr}$ followed by a 
 quiescent period with only sporadic star formation; b) LSBs'
discs are not young. The
age of the LSB disc galaxies for which colors have been 
measured are all larger than about 10 ${\rm Gyr}$.

\end{abstract}

\section{Introduction}

Late-type Low Surface Brightness galaxies (LSBs) are considered to be very young
stellar systems, because of their rather blue colors (de Blok, van der Hulst \& 
Bothun 1995, McGaugh \& Bothun 1996) and very low oxygen abundances 
(McGaugh, 1994). Based on these observational evidences there
have been recently theoretical suggestions that LSBs are formed 
inside dark matter halos that collapsed very recently, at 
$z\le 1$, from density fluctuations of small amplitude 
(Dalcanton, Spergel, \& Summers 1996, Mo, McGaugh, \& Bothun 
1994). 

In this work we study the colors of LSBs from the point of view 
of synthetic stellar populations (SSP), and show that LSBs
could not be as young as claimed in the quoted literature.
Recently one of us (PP) has obtained a stellar Initial Mass
Function (hereafter P-IMF) starting from high-resolution numerical
simulations of the supersonic random motions in the interstellar 
medium (Nordlund \& Padoan, 1997; Padoan, Jones \& 
Nordlund,1997).
Here we will plug this P-IMF into the latest version of our 
synthetic 
stellar population code which is based on Jimenez \& MacDonald (1997)  
evolutionary tracks
 and Kurucz atmospheric models (Kurucz 1992). With this we
compute synthetic colors and colors gradients for LSBs
(section 2) and we show how these can be used to set tight
bounds on the ages of their stellar discs (section 3). We also
show that the color gradients are well fitted (section 4), and
we speculate on the cosmological implications of these results
in section 5.

\section{Synthetic stellar populations for LSBs}

In the following when we will refer to LSBs' we will always mean 
the sample of late-type disc galaxies observed by de Blok, 
van der Hulst \& Bothun (1995). For each galaxy of their sample 
the HI surface density, and the surface brightness profiles in 
several bands are published. 

LSBs are found to be rather blue; the color tends to become bluer
in the outer regions of their discs. De Blok, van der Hulst 
\& Bothun (1995) noted that it is difficult to understand the 
colors of LSBs, 
if their stellar population is old or forming at a declining rate. 
McGaugh and Bothun (1996) from the analysis of their sample
concluded that the stellar 
populations in LSBs must be very young, because of the very blue colors 
and of the very low metallicity. In fact an IMF appropriate to the 
solar neighbourhood, like the one by
Miller and Scalo (1979), has a shape very flat for ${\rm M}\leq 0.1 {\rm
M}_{\odot}$ and this results in too red V-I colors when B-V are properly
fitted.

Since the discs of LSBs are rather quiescent when compared with 
HSB discs, we suppose that their colors are an excellent probe 
of their stellar IMF. Although this can at most be taken as
first approximation, it gives an excellent fit to many
observed relations, as we will show. Moreover, it allows us to
probe to which extent our P-IMF can provide a
realistic interpretation of observed data. At variance with
other IMF, in the P-IMF there are no free
parameters, and it is 
based on a model for the structure and dynamics of molecular 
clouds, that has
strong observational support (Padoan, Jones, \& Nordlund 1997, 
Padoan \& Nordlund 1997).  

The P-IMF is designed to model large scale star formation, and 
contains a
dependence on mean density $n$, temperature $T$, and velocity 
dispersion
$\sigma_{v}$ of the star forming gas. The mean stellar mass is 
given by:

\begin{equation}
M_{*}=1\rm {M}_{\odot}\left(\frac{T}{10\,K}\right)^2\left(\frac{n}{10\, cm^{-3}}\right)^{-1/2}
\left(\frac{\sigma_{v}}{5\, km/s}\right)^{-1}
\label{eq1}
\end{equation}

As a significant example we apply the P-IMF to a simple 
exponential disc model, with height-scale  
equal to $100\rm\, {pc}$, length scale equal to $3\, \rm{Kpc}$, 
and total mass
equal to $\rm{M_{D}}=3\times10^9 \rm {M}_{\odot}$, a set of
parameters chosen to be representative of the LSBs. Our
results
about colors depend only slightly on these
particular values, however.

As a measure of the gas velocity dispersion we use the disc vertical
velocity dispersion. We also assume that all stars are formed 
in a cold gas phase, at T$=10\, K$.

Note that the same stellar mass would be obtained if the vertical velocity
dispersion, instead of the height-scale, were kept constant along the radius,
because of the dependence on velocity dispersion and density in 
equation (1).

Fig.~1 shows the IMF predicted for such a disc at 1$ \rm {kpc}$ 
and
6$ \rm {kpc}$ from its center. The IMF is more massive 
than the Miller-Scalo (dashed line), but also less broad. The 
IMF at 
6$ \rm {kpc}$ is also more massive than at 1$ \rm {kpc}$. 
We then expect that with these properties the stellar populations
which will form will be rather blue, and will become bluer at
larger distances from the center, as is observed in LSBs.

To compute the synthetic colors we used the latest 
version of our synthetic stellar population code (Jimenez et al. 1996). 
The code uses the library of stellar tracks computed  with 
JMSTAR9 
 and the set of atmospheric models calculated 
by Kurucz (Kurucz 1992). A careful treatment of {\rm all} 
evolutionary stages 
has been done following the prescriptions in Jimenez et al. (1995), and Jimenez 
et al. (1996). Different star formation rates and stellar IMF 
are incorporated in the code, so a large parameter space can be investigated.

We find that the star formation in LSBs can be
adequately described with 
an initial burst, followed by a quiescent evolution up to the 
present time. It has been already remarked (van der Hulst et al.,
1993) that LSBs' gas surface
densities are too low to allow efficient star formation according to
Kennicut criterion (Kennicut 1989). Therefore it is reasonable to
argue that significant star formation is limited to an initial
burst. The duration of the burst is almost irrelevant 
to the colors, because of its rather old age, but 
it cannot be much longer than a few $10^7$ yr, in order to be 
consistent with the low metallicity of the synthetic stellar 
population, and with the low oxygen abundance of the HII 
regions observed by McGaugh (1994) in LSBs.

We find that the colors of LSBs are not difficult to reproduce, 
as long as stars smaller than $1 \rm{M}_{\odot}$ are not 
as numerous as in the solar-neighborhood population, which would 
give a 
too red V-I color, and as long as a low metallicity is used. 
Indeed, one can easily see, 
from the theoretical models by Kurucz (1992), that even a 
{\it single} star with low 
metallicity (Z=0.0002) can reproduce the colors of LSBs. As an 
example, the colors
of a typical galaxy from the sample of de Blok, van der Hulst, 
\& Bothun, namely F568-V1,
are: U-B=-0.16, B-V=0.57, B-R=0.91, V-I=0.77 (luminosity 
weighted); the colors
of a Kurucz model with temperature T=5500 K, $\log$(g)=4.5, 
Z=0.0002
are: U-B=-0.17, B-V=0.56, B-R=0.94, V-I=0.75. This model 
corresponds to
a star of $0.94 \rm{M}_{\odot}$, having a lifetime of 11 Gyr. 
Obviously, the reason for 
such a good match does not lie in the fact that the stellar IMF does not 
contain any star more
massive than $1 \rm{M}_{\odot}$, as suggested in the past 
(Romanishin, Strom, \& Strom 1983, Schombert et al. 1990), but 
simply in the fact that $0.94 \rm{M}_{\odot}$ is the mass 
at the turn-off for the stellar population of F568-V1 in our
model, which gives an age for this galaxy's disc of about 11
Gyr.

\section{The age of LSBs}

In Fig.~2 we plot the time evolution of the colors for a very low 
metallicity ($Z=0.0002$), and in Fig.~3 for a higher metallicity 
($Z=0.0040$).

In order to compare the theoretical prediction with the observed colors,
we have used the mean values of the luminosity-weighted colors listed
in Table~4 of de Block, van der Hulst, \& Bothun (1995).
Since the color of a stellar population is affected by age and
metallicity, we also plot in Fig.~2 the mean of the observed colors,
excluding the three galaxies for which U-B is observed and has a positive 
value. The error bars represent the dispersion around the mean. It is 
clear 
that the fit is excellent for an age of $12 \,\rm{Gyr}$, and that an age
$\le 9 \,\rm{Gyr}$ is definitely inconsistent with the data.

In
Fig.~3 we plot the mean of the colors for the three galaxies with
positive U-B. These redder galaxies are better fitted by a higher 
metallicity, 
$Z=0.0040$, which is one fifth of the solar metallicity, and is one of 
the highest metallicity estimated by McGaugh (1994) in LSB HII regions. 
The best fit for the age is $9\, \rm{Gyr}$.   

The effect of the metallicity on the colors is illustrated
in Fig.~4 and 5, where we show the trajectories of the time evolution of our 
models in color-color diagrams. It is evident that we do not find LSBs 
younger than $9\,{\rm Gyr}$, for any metallicity consistent with the
observations. The spread in colors is a result of the spread in
metallicity, as is shown by the remarkable agreement between the
trajectories and the observations. For instance, in the (B-V,U-B)
diagram, where the trajectories are well separated, also the observed
points show a similar spread in U-B. On the other hand, in the
(B-R,B-V) diagram, where the theoretical trajectories are almost coincident,
also the observational points are nicely aligned around the
trajectories.

Therefore {\it the ages of LSBs' discs rule out the possibility 
that they 
formed from primordial density fluctuations of low amplitude, 
collapsed at $z\le1$}. Such old ages may seem difficult to
reconcile with those of the relatively young stellar populations in
normal late-type galaxies, that have U-B and B-V colors comparable
to those of LSBs, and B-R and V-I even redder. However, the very blue
U-B and B-V colors in LSBs are very well explained by the very low
metallicities, rather than by the young stellar ages, and the B-R and V-I
colors are explained by the lack of small stars (as the P-IMF predicts), 
in comparison with a Miller-Scalo
IMF.

The diagram (B-V,U-B) shown in Fig.~5 is particularly important, 
because it can be used to estimate the age of single galaxies, without
an independent determination of the metallicity of its stellar population.
In fact, in that diagram the time evolution is almost horizontal, along
B-V, while the metallicity variations are almost vertical, along U-B.
In other words, the degeneracy age-metallicity in the colors is broken
in such a digram. We can therefore see that galaxies of different 
metallicities have all about the same age (11-12 Gyr). The horizontal
dispersion of the observational points, along B-V, is approximately 
0.1 mag, which is comparable to the observational uncertainty. Therefore,
the determination of the age of LSB discs with presently available
photometry, and without an independent estimate of the metallicity,
has an uncertainty of $\pm 2.0$ Gyr (0.1 mag in B-V).

\section{Color gradients}

An interesting feature of LSBs is their color gradient: LSBs are bluer in the 
periphery than near the center of their disc (de Blok, van der Hulst, \& 
Bothun 1995). 

Our theoretical models predict a color gradient in agreement with the
observations. In fact, the exponential disc model has a volume density 
that decreases with increasing radius, and equation (1) shows that
the typical stellar mass in the IMF grows with decreasing gas density, 
producing increasingly bluer colors.

We have computed the color gradients per disc length-scale, for a 
model with an age of 12 Gyr, and metallicity Z=0.0002. In Table~1
we show the results, compared with the observational data, which are 
obtained from the mean of the values listed by de Blok, van der Hulst, 
\& Bothun (1995), in their Table~3. Again, we have excluded from the mean the 
three galaxies with U-B$>0$, since they require a metallicity significantly larger
than Z=0.0002. Together with the mean gradients, we give the mean of the 
errors listed by the above mentioned authors.

The agreement between observational data and theory is striking. Note
that the model is just the one that best fits the colors of LSBs, as shown
in Fig.~2, rather than being an ad hoc model which fits the 
color gradients.
Therefore {\it the color gradient of LSBs indicates that the stellar IMF
is more massive towards the periphery of the discs than near the 
center}, as predicted by our P-IMF.

\section{Conclusions}

In this work we have shown that the P-IMF, applied to a simple
exponential disc model, allows an excellent description of the colors
and color gradients of LSBs. This allows us to draw a few 
interesting consequences:

\begin{itemize}

\item The Miller-Scalo IMF produces too red V-I colors, and therefore 
cannot describe the stellar population of LSB galaxies;

\item The P-IMF, applied to a simple exponential disc model with an
initial burst of star formation, produces
excellent fits of the LSBs' colors and color gradients;

\item The metallicity of LSB stellar populations ranges from practically
zero to about one fifth solar. 

\item Although most stars in LSBs are formed in an initial burst, a relation between
colors and surface brightness is not expected, because the colors are strongly
affected also by the metallicity. 

\item The age of LSBs, inferred from the UBVRI colors, is between 
$9$ and $13\, \rm{Gyr}$. These disc populations are therefore about as 
old as the disc of our Galaxy.

\item Since LSBs galaxies are old they cannot be explained as late collapsed objects (low density fluctuations at $z\le1$), therefore their origin remains still unexplained.

\end{itemize}

\acknowledgements

This work has been supported by the Danish National Research Foundation
through its establishment of the Theoretical Astrophysics Center.
RJ and VAD thank TAC for the kind hospitality and
support.

\clearpage

{\bf Figure Captions}
\noindent

{\bf Figure 1}: The theoretical IMF at 1 Kpc and 6 Kpc from the center of the disc
 (continuous lines). The dashed line shows the Miller-Scalo IMF. The theoretical 
 IMF are more massive than the Miller-Scalo one.

{\bf Figure 2}: The plot shows the time evolution of the colors in a model with metallicity $Z=0.0002$ and star formation in a initial burst of $5\times10^7$ yr -- 
the 
continuous line 
is U-B, the dotted line is B-V, the dashed-dot line is V-I and 
the dashed line is B-R. The symbols represent the observed mean values for the 
sample of LSBs (de Blok et al. 1995) excluding the galaxies with U-B $>$ $0$.

{\bf Figure 3}: The same as Fig. 2 but for Z=0.0040. In this case the symbols 
represent only the mean value for the three galaxies in de Blok et al. (1995) 
sample with U-B $>$ $0$.

{\bf Figure 4}:
Trajectories of the time evolution of the models in the
(B-R,B-V)
diagram. The continuous line is the model with $Z=0.0002$, and the dashed
line the
model with $Z=0.0040$. The diamonds are the observed luminosity weighted colors
of LSB discs, from de Block et al. (1995). 
The trajectories are from 1 Gyr (left)
 to
14 Gyr (right). 
The vertical line marks the 9 Gyr age for the lower metallicity.
 Here the
degeneracy is not broken: age and metallicity change in almost 
the same direction,
and in fact the observational points are now nicely aligned 
with the theoretical
 trajectories.
The vertical and the horizontal lines
mark the age of 9 Gyr for the lower metallicity. On the right
of the vertical line and above the horizontal one, the models are
older than 9 Gyr.

{\bf Figure 5}: 
The same for the (B-V, U-B) diagram. 
All galaxies are clearly older than 9 Gyr. This diagram breaks the degeneracy
age-metallicity in the colors.

\clearpage

\begin{table}
\begin{tabular}{|c|c|c|c|c|}
\hline
 & $\Delta$ (U-B) & $\Delta$ (B-V) & $\Delta$ (B-R) & $\Delta$ (V-I) \\
\hline\hline
de Blok et al. & $-0.15 \pm 0.14$ & $ -0.07 \pm 0.12$ & $-0.18 \pm 0.08 $ & $-0.18 \pm 0.17$ \\
this work & -0.04 & -0.07 & -0.14 & -0.15 \\
\hline
\end{tabular}
\caption{}
\end{table}

\clearpage
\begin{figure}
\centering
\leavevmode
\epsfxsize=1.0
\columnwidth
\epsfbox{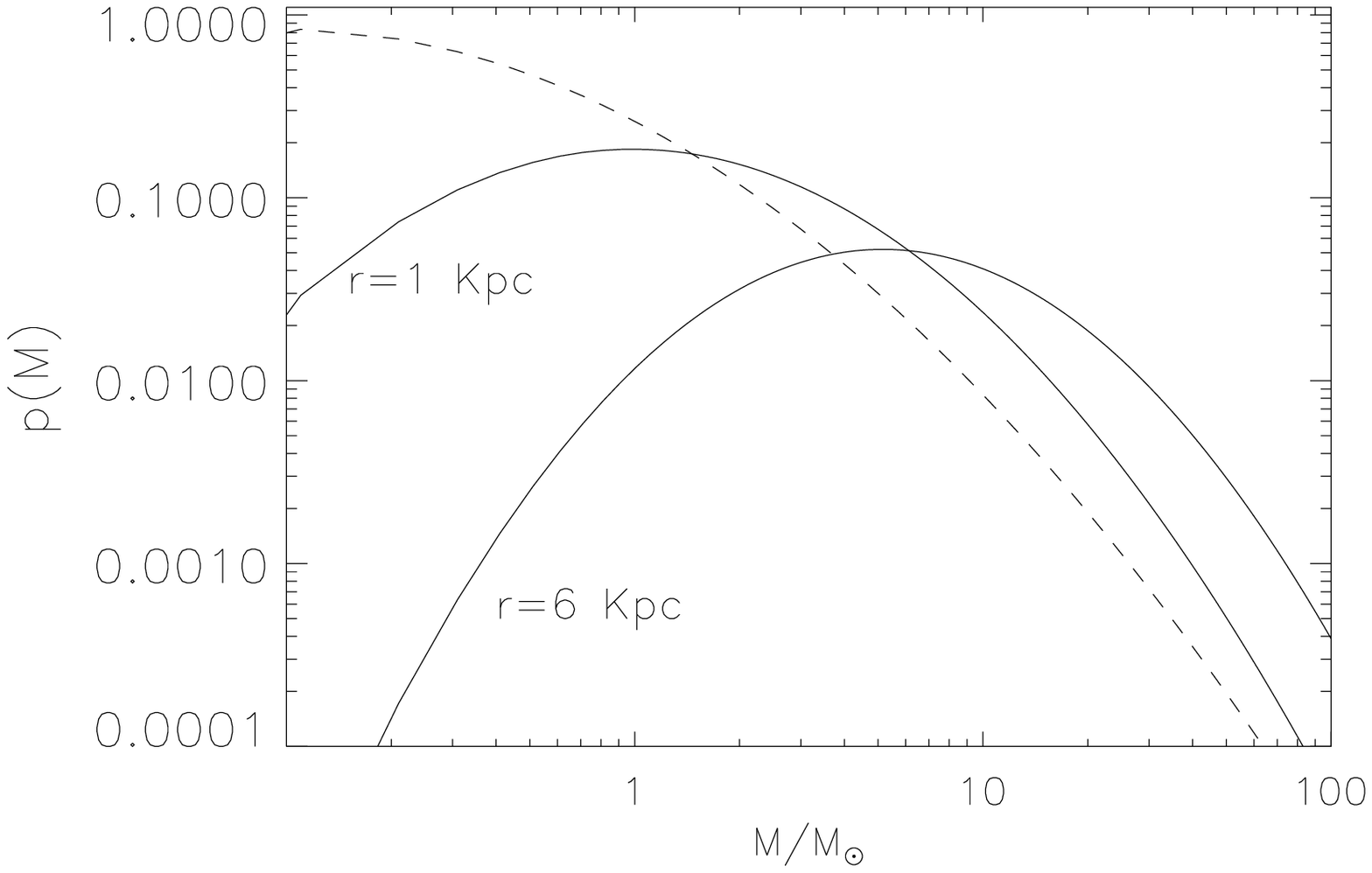}
\caption[]{}
\end{figure}
\clearpage
\begin{figure}
\centering
\leavevmode
\epsfxsize=1.0
\columnwidth
\epsfbox{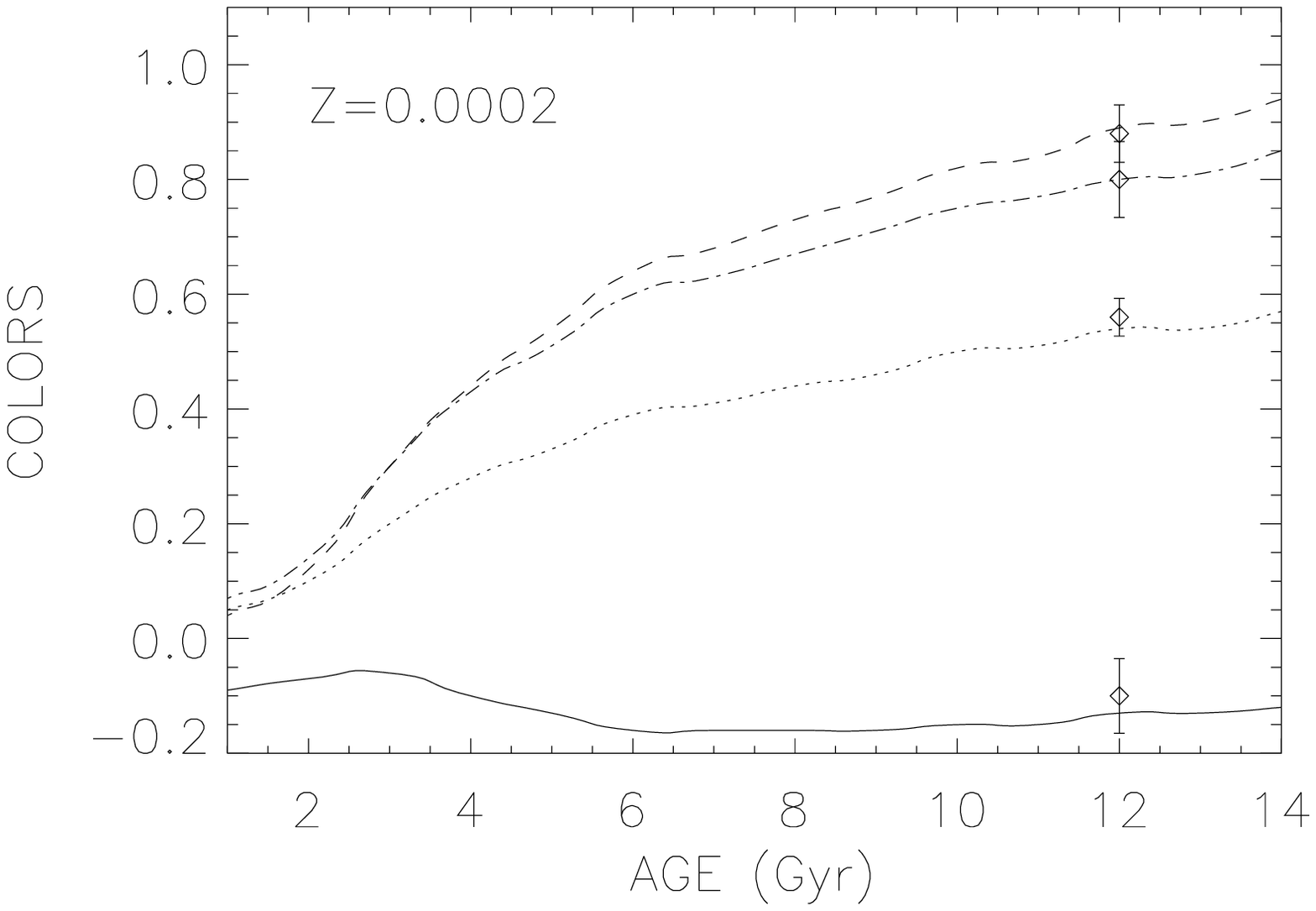}
\caption[]{}
\end{figure}
\clearpage
\begin{figure}
\centering
\leavevmode
\epsfxsize=1.0
\columnwidth
\epsfbox{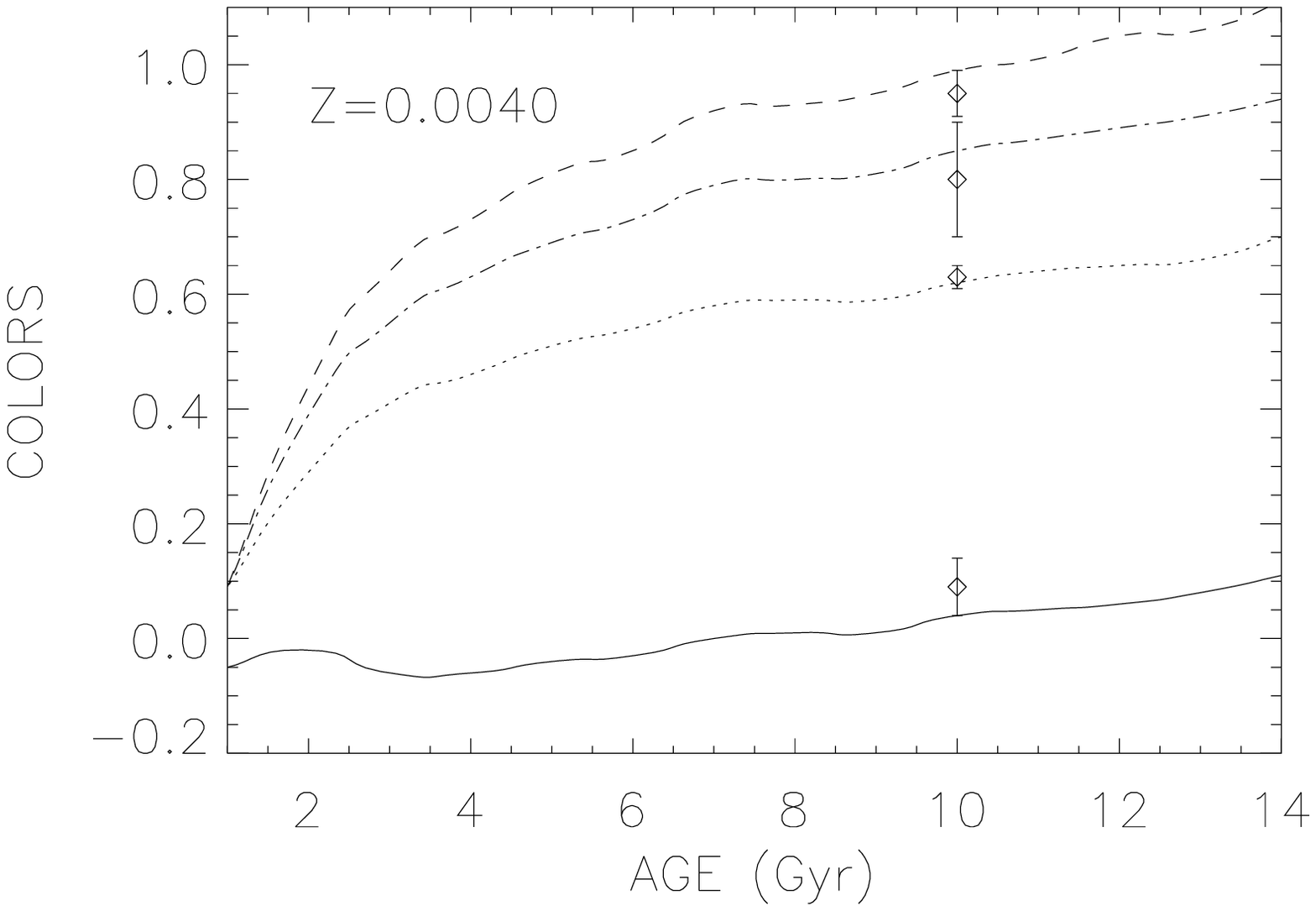}
\caption[]{}
\end{figure}
\clearpage
\begin{figure}
\centering
\leavevmode
\epsfxsize=1.0
\columnwidth
\epsfbox{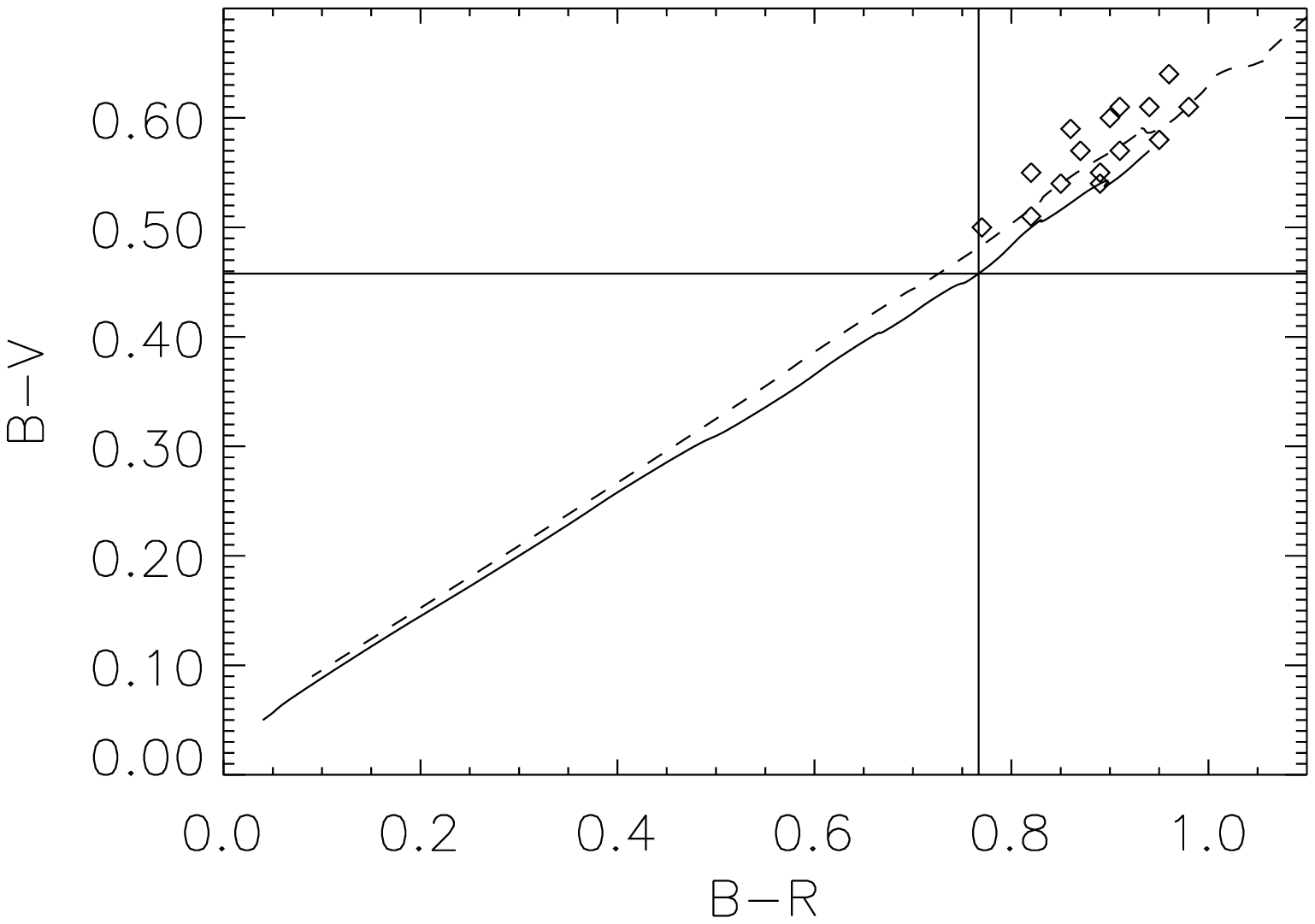}
\caption[]{}
\end{figure}
\clearpage
\begin{figure}
\centering
\leavevmode
\epsfxsize=1.0
\columnwidth
\epsfbox{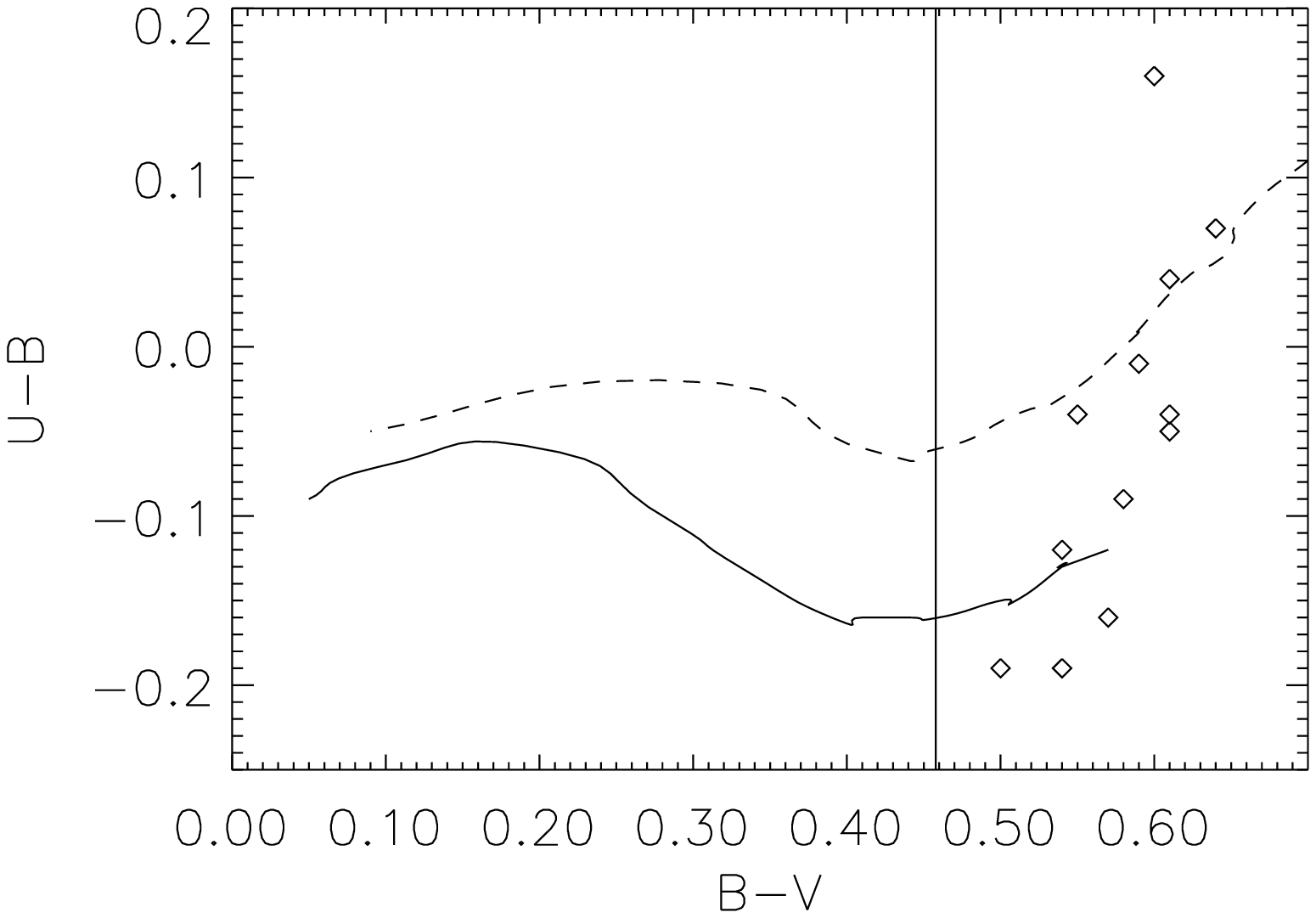}
\caption[]{}
\end{figure}

\end{document}